\documentstyle[preprint,floats,prd,aps]{revtex}

\begin{document}
\preprint{\vbox{\hbox{ORNL-CTP-96-04 \hfill}
                \hbox{UH 511-849-96 \hfill}
                \hbox{May 1, 1996 \hfill}}}

\def\ra{\rightarrow}
\def\be{\begin{equation}}
\def\ee{\end{equation}}

\title{Prospects for detecting an $\eta_c'$ \\
in two photon processes}

 \author{T. Barnes \\
        Physics Division, Oak Ridge National Laboratory, \\ Oak Ridge,
                             TN 37831-6373 and \\ 
Department of Physics, University of Tennessee, \\
        Knoxville, TN 37996-1501
\bigskip \\
              T.E. Browder and S.F. Tuan \\
     Department of Physics, University of Hawaii at Manoa, \\
       Honolulu, HI 96822}

\maketitle
\begin{abstract}
We argue that an experimental search for an
$\eta_c'$, the first radial excitation of the
$\eta_c(2980)$, may be carried out using 
the two photon process
$e^+e^- \ra e^+e^- \gamma \gamma \ra e^+e^-\eta_c'$. 
We estimate the partial width
$\Gamma_{\gamma \gamma}(\eta_c')$ and
the branching fraction $B(\eta_c' \ra h)$, where $h$ is an exclusive hadronic channel, 
and find that for
$h = K^o_s K^\pm \pi^\mp$ it may be possible to observe this state
in two photon collisions at CLEO-II.
\end{abstract}

The $\eta_c'$ meson is
 the first radial excitation of the $\eta_c(2980)$, the
charmonium  ground state with J$^{\rm PC} = 0^{-+}$.
The mass and couplings of this state
can be used to test potential models and lattice QCD calculations.
In this paper we discuss possible experimental signatures and event rates
for the
$\eta_c'$, and note that it may be observable in future CLEO-II data
sets.

The search for 
the $\eta_c'$ \cite{crystal}, expected near a mass of $3600$ MeV, 
has intrigued workers in the field for over a decade. In an
$e^+e^-$ experiment the Crystal Ball collaboration \cite{crystal} 
reported evidence for this state near 3590~MeV, with a branching fraction of
$0.2\% < B(\psi' \ra \gamma + \eta_c')<1.3\% $
at 95\% confidence level.
However the Crystal Ball observation 
has not been 
confirmed by subsequent experiments over the intervening 13 
years. 

Since the relative rates for competing 
electromagnetic decays and 
transitions to lower charmonium 
states are expected to be small for the $\eta_c'$, 
the relation \cite{trig_tuan2}
\begin{equation}
             B(\eta_c' \ra h) \cong B(\eta_c \ra h)\ ,               
\end{equation}
where $h$ is an exclusive final state,
can be assumed as a working hypothesis.
Since the $\eta_c(2980)$ branching fractions are known for many
hadronic channels \cite{pdg} we can then estimate event rates for $\eta_c^{'}$ 
decays into the same modes.

Of course there are important 
uncertainties in these estimates;
it is well known that the analogue of Eq.(1) for the $\psi$ and $\psi'$ 
does not account for the puzzling weakness 
of certain $\psi^{'}\to h$ modes, notably vector+pseudoscalar 
final states such as $h=\rho \pi$ and $ K^{*} \bar{K}+h.c.$ \cite{pdg}. 
In part these small $\psi'$ branching fractions simply 
reflect the presence of
important $\psi' \to 
\psi + h$ ``hadronic cascade'' modes, 
which have a large branching fraction of $57(4)\% $
\cite{pdg}. 
We expect these
cascade modes to be much less important
for the $\eta_c'$,
due to the larger 
annihilation width of a $0^{-+}$ state.
Of course there remains considerable uncertainty regarding these branching
fractions, which merit careful investigation when more data on radial $c\bar c$
states becomes available.

For 
a three-body or four-body hadronic decay mode $h$,
such as those considered here for the $\eta_c^{'}$, the 
corresponding QCD-motivated
relation 
$B(\psi^{'}\to h)/ B(\psi\to h)\approx 0.13$,
discussed by Brodsky et al. \cite{trig_tuan2} 
appears to be reasonably well satisfied \cite{pdg},
and the recent BES measurement of
$B(\psi^{'}\to K^+ K^-\pi^+\pi^-)$ is also consistent with this estimate
\cite{gu}.

As a first possibility for observing the $\eta_c'$ we 
estimate the number of events expected at BES in the 
decay chain $\psi' \ra \gamma \eta_c' \ra \gamma h$,
assuming that Eq.(1) is approximately correct. 
At present there are about $3.4 \times 10^6 \psi'$ events in the BES 
data set, and the two largest $\eta_c(2980)$
branching fractions 
are to $K^o_s K^\pm \pi^\mp 
(\cong 1.5 \pm 0.4\%)$ and $\pi^+\pi^- K^+K^- (\cong 2^{+0.7}_{-0.6})\%$. 
Taking the lower end of the range reported  
by Crystal Ball \cite{crystal}, $B(\psi' \ra \eta_c'+ \gamma
) = 0.2\%$, we would expect 102 events from the decay
$\psi' \ra  \gamma \eta_c' \ra \gamma \ h$, 
$\eta_c' \ra K^o_sK^\pm \pi^\mp$, and 136 events from the corresponding decay 
with $\eta_c' \ra \pi^+\pi^-K^+K^-$.
The average efficiency for detection is approximately 5\% with the 
current BES calorimeter \cite{gu} 
(the photon produced in the radiative transition 
$\psi' \to \gamma \eta_c'$ 
is soft, with an energy of $< 90$ MeV), so unfortunately the
total expected number of events is reduced to
$5.1 \pm 1.4$ 
and $6.8^{+2.4}_{-2.0}$
respectively. In a missing-$\gamma$ analysis, in which 
one does not detect 
the low-energy $\gamma$, the efficiencies
are estimated to be 15 - 25\% \cite{olsen}, so the number of events expected
would be
$15.3 \pm 4.2$
and $20.4^{+7.2}_{-6.0}$ (for 15\%) and 
$25.5 \pm 7.0$ and $34.0^{+12.0}_{-10.0}$ (for 25\%).
In view of the small event sample anticipated, 
success of this approach may require a
BES upgrade to a detector similar to the Crystal Barrel or a larger
integrated luminosity.

Although the transition
$\psi' \ra \gamma\eta_c'$ required to detect the
$\eta_c'$ in this approach has 
not been observed since the 
Crystal Ball experiment \cite{crystal}, we note that the 
branching fraction of $0.2 \% $ we used for our estimates is 
quite conservative; in comparison,
Godfrey and Isgur \cite{GI}
quote a $\langle \psi'|\mu | \eta_c'\rangle$ 
transition magnetic moment that corresponds to 
$B(\psi' \to \gamma\eta_c'(3590)) = 0.8 \% $. Of course this branching fraction
could be somewhat smaller; it scales as
$E_\gamma^3$ and roughly as $1/m_c^2$, so a somewhat higher 
$\eta_c'$ mass and a larger $m_c$ value than is normally assumed could 
conceivably reduce the
branching fraction to $0.2 \% $.

E835 proposes to search for the $\eta_c'$ at FERMILAB in
the process $p\bar{p} \ra
\eta_c' \ra \gamma \gamma$, which requires rejection of a large hadronic
background.
The preceding experiment, E760, searched for evidence of an $\eta_c'$ 
in the mass range 3584-3624~MeV
without success. For the $\eta_c$, E760 reported 
$B(\eta_c \ra  \gamma \gamma)$ 
$=(2.80^{+0.67}_{-0.58} \pm 1.0) \times 10^{-4}$ \cite{E760}; this
is close to a perturbative QCD estimate of 
$B(\eta_c \ra \gamma \gamma) = 3.0 \times 10^{-4}$
\cite{cester},
but the signal is seen
at a mass of
$2988^{+3.3}_{-3.1}$ MeV, somewhat higher than the PDG value of 
$2978\pm 1.9$~MeV.
It is especially interesting that E760 finds a rather large width for the
$\eta_c$,
$\Gamma_{\eta_c}=23.9^{+13.6}_{-7.1}$~MeV,
compared to the PDG value of $10.3^{+3.8}_{-3.4}$~MeV. This number is required
to convert the E760 $\eta_c\to\gamma\gamma$ branching fraction to the theoretically
more accessible $\gamma\gamma$ 
partial width, so until $\Gamma_{\eta_c}$ is accurately measured
a definitive comparison with theory will not be
possible.

Of course the absence of an $\eta_c'(3600)$ signal at 
E760 may be due to a
weak $\eta_c^{'}-p\bar p$ coupling relative 
to $\eta_c-p\bar p$, as has been observed
in the $\psi^{'}-p\bar p$ and $\psi-p\bar p$ couplings; the PDG branching
fraction $B(\psi^{'}\to p\bar p)$ is an order of magnitude smaller than
$B(\psi\to p\bar p)$.
Although Brodsky and Lepage \cite{brodsky} anticipated a weak coupling of
the $\eta_c^{'}$ to $p \bar{p}$ in perturbative QCD, their results should apply
to the ground state $\eta_c$ as well; the expected weak $\eta_c-p \bar{p}$ coupling is not
supported by experiment.
Clearly the relative annihilation amplitudes of ground-state and
radial quarkonia 
to exclusive hadronic final states is a vitally important question for these
experiments, and it is unfortunately not well understood at present.

From Eq.(1), we expect the $\eta_c'$ branching 
fraction ratio $B(\eta_c'\to \gamma\gamma )$  $ / $ $ B(\eta_c'\to h)$ 
to be comparable to the $\eta_c(2980)$ ratio.
Thus in the absence of a magnetic spectrometer to detect
exclusive hadronic modes $h$, $p \bar{p}$ experiments such as
E835 will have to contend with a correspondingly weak signal
in the search for the $\eta_c'$.

Having discussed searches for the $\eta_c'$ in $\psi'\to\gamma\eta_c'$ and 
$p\bar p\to\eta_c'$, we now consider a third possibility,
the reaction 
$$e^+e^- \ra e^+e^-\gamma \gamma \ra 
e^+e^-\eta_c' \ . $$ 
It is interesting to consider whether 
this process is likely to lead to an observable signal with present
and future data sets. (One of us previously suggested that the
$\chi'_{cJ}$ (J$=0,2$) states
expected near 3.95 GeV, 
the radial excitations of the $\chi_{cJ}$ 
states, could be observed 
in this manner \cite{tuan1}.)

CLEO-II and LEP experiments such as L3 are
well equipped to study exclusive hadronic modes $h$.
Indeed, these experiments 
have confirmed the existence of the ground state $\eta_c(2980)$ 
in this manner \cite{fulton,savinov,L3}. To estimate the number of
$\eta_c'$ events in these $\gamma\gamma$ experiments
we again assume the validity Eq.(1); in addition we require a theoretical
estimate of the $\Gamma_{\gamma \gamma}(\eta_c')$ partial width.

The partial width $\Gamma_{\gamma \gamma}(\eta_c'(3590))$ 
has been calculated
by Ackleh and Barnes \cite{ackbarnes} using 
nonrelativistic quark potential model wavefunctions 
attached to relativistic Feynman decay amplitudes; this gives 
$\Gamma_{\gamma \gamma}(\eta_c'(3590)) = 3.7$~keV. For comparison, this approach
applied to
the $\eta_c$ gave the prediction
\begin{equation}
\Gamma^{thy.}_{\gamma \gamma}(\eta_c) =  4.8\ {\rm keV}  \ ,
\end{equation}
which
is
consistent with recent E760, 
CLEO-II 
and L3 measurements; 
\begin{equation}
\Gamma^{expt.}_{\gamma\gamma}(\eta_c) = 
\left\{
\begin{array}{ll}
6.7 {+2.4 \atop -1.7} \pm 2.3  \; {\rm keV}
\ &   {\rm E760} \ \cite{E760}   \\
4.3 \pm 1.0 \pm 0.7 \pm 1.4 \; {\rm keV}
\ &   {\rm CLEO} \ \cite{savinov}   \\
8.0 \pm 2.3 \pm 2.4 \; {\rm keV}
\ &   {\rm L3} \ \cite{L3}    \  . \\ 
\end{array}
\right.
\end{equation}
\noindent
(The E760 reference gives a more complete summary of previous 
experimental results, in their Table V.)

An important conclusion of Ackleh and Barnes is that
$\Gamma_{\gamma \gamma}$ widths of quarkonia are not strongly 
suppressed with radial excitation for light {\it or} heavy quarks.
This suggests that a radially excited $\eta_c'(\approx 3600)$ may 
be observable in $\gamma \gamma$ collisions, provided that the exclusive 
hadronic channel studied does not have an unusually weak 
coupling to the $\eta_c'$.

The level of uncertainty in the 
$\eta_c'$ $\gamma \gamma$ width 
in the model of Ackleh and Barnes can be estimated by
considering
a likely range of 
$M(\eta_c')$ values, and
allowing plausible variations in the three model
parameters.
The effect of varying $M(\eta_c^{'})$ over the
range $3590\pm 50$~MeV is to change $\Gamma_{\gamma\gamma}(\eta_c^{'})$ by
$\pm 0.15$~keV, a $\pm 4\% $ change. Variations in the
three parameters $\alpha_s$, $b$ (GeV/fm), 
and $m_c$ have also been considered.
The effect of varying the first two is quite small;
the width is most sensitive to the charm quark mass
$m_c$, which was taken to 
be $1.4$ GeV in Ref.~\cite{ackbarnes} and our Eq.(2). As we increase $m_c$ from
$1.3$ to $1.8$ GeV, 
$\Gamma_{\gamma\gamma}(\eta_c^{'})$ decreases monotonically by
about a factor of two. Fortunately, $\Gamma_{\gamma\gamma}(\eta_c^{'})$ 
and $\Gamma_{\gamma\gamma}(\eta_c)$ show similar $m_c$ dependences, 
so the ratio 
$\Gamma_{\gamma\gamma}(\eta_c^{'})/ \Gamma_{\gamma\gamma}(\eta_c)$
only changes from $0.73$ to $0.77$. Comparison with the experimental
$\eta_c$ values then supports
our use of 
$\Gamma_{\gamma\gamma}(\eta_c^{'})=3.7$~keV as a reasonable first estimate
in determining numbers of events.
It should be noted however that no radial 
excitations have yet been identified in $\gamma \gamma$ collisions, so 
this width calculation is a theoretical estimate in
a regime in which the theory has not been tested.

In two photon physics the number of events
is related to the production cross section in $e^+e^-$ by  
\cite{fulton}
\be
N_{obs} = \epsilon \; L \; B(R_{c\bar{c}} \ra F_s) \;  
\sigma_{exp}(e^+e^- \ra e^+e^-R_{c\bar{c}} )   \ ,
\ee
where $\epsilon$ is the efficiency for detecting the final state $F_s$,
$B(R_{c\bar{c}} \ra F_s)$ is the branching fraction for 
$R_{c\bar{c}}$ into $F_s$, $L$ is
the luminosity, and $N_{obs}$ is the number of $R_{c\bar{c}} 
\ra F_s$ events observed. Here we specialize to 
$R_{c\bar{c}} = \eta_c(2980), R'_{c\bar{c}} = \eta_c'(3590),
F_s$
an exclusive final state $h$, and the luminosities are $L$ and $L'$ 
respectively.
The ratio of the number of events $N'_{obs}$ expected 
for an $\eta_c'(3590)$ to the
$\eta_c(2980)$ number is then
\be
{N'_{obs} \over N_{obs} } = 
{\epsilon' L'
\over
\epsilon L } \ 
{
B(\eta'_c \ra F_s) 
\over
B(\eta_c \ra F_s) } \ 
{
\Gamma_{\gamma\gamma}(\eta'_c) 
\over 
\Gamma_{\gamma\gamma}(\eta_c) 
} \ ,
\ee
where we have taken the ratio of $\gamma \gamma$ 
production cross sections for an $\eta_c'$ and the $\eta_c(2980)$ 
to be the ratio of their respective $\gamma\gamma$ widths, up to photon 
flux factors \cite{cahn}. 

For the photon
flux factors we have $L'/L = 1/2.02$ \cite{EPA}. 
Since we assume Eq.(1) it follows that 
$\epsilon^{'} = \epsilon$,  
i.e. the detection efficiencies are equal because the final
states are identical. 
(We assume that the difference in invariant mass of the two states
has no significant effect on the efficiency.) 
From the theoretical estimate above we have
$\Gamma_{\gamma\gamma}(\eta_c')/\Gamma_{\gamma\gamma}(\eta_c) 
\approx  1/1.3$. Using the yield
$N_{obs}$ $=54.1 \pm 12.6$ events 
for $h = K^o_s K^ \pm \pi^\mp$ reported by CLEO-II 
for the $\eta_c$, 
(c.f. Savinov and Fulton in ref. \cite{savinov}), 
we obtain $N^{'}_{obs} \cong 20.6 \pm 4.8$ for the $\eta_c^{'}$. 
Note the large number of exclusive
events expected at CLEO-II, in contrast with LEP experiment
L3, for which the corresponding $N_{obs}$ for
$\eta_c \ra K^o_s K^\pm  \pi^\mp$ is 3 events \cite{L3}. 
The difference is primarily due to the 
smaller integrated luminosity available at the
LEP experiments.

It would be 
useful for CLEO-II
to extend their search to include $\eta_c' \ra \pi^+\pi^-K^+K^-$,
which is also expected to have a large branching ratio ($\sim 2$\% according 
to Eq.(1)). Of course
the CLEO-II detector (now at the CESR $e^+e^-$ storage ring, and running at a
center-of-mass energy near the $\Upsilon (4S)$, $\approx 10.6$ GeV),  
is continually accumulating data, so prospects for finding
the $\eta_c'$ will improve with time. 
The value of $N^{'}_{obs}$ estimated above assumes a 
dataset with an integrated luminosity of
3.0 fb$^{-1}$, 
and should be scaled with the size of the dataset.
It is anticipated for example that the
CLEO dataset will have an
integrated luminosity
of about 10.0 fb$^{-1}$ 
by 1998 \cite{Berkelman}, so our expected number of 
$\eta_c'$ events would then be $N'_{obs} = 68.7\pm 16.0$.

In summary,
we have considered some possible experimental signatures for
$\eta_c^{'}$ production, and suggest that 
$e^+e^- \ra e^+e^-\gamma \gamma \ra e^+e^-\eta_c'$,
$\eta_c'\to K_s^0 K^{\pm} \pi^{\mp}$  
may be feasible at CLEO-II.
Searches for the $\eta_c^{'}$ in this process can be conducted 
at CLEO-II, LEP, and future B factories.

One of us (S.F.T.) wishes to thank Dr. Sergei Sadovsky for a useful
comment. This work was supported in part by the US Department of Energy under
Grant DE-FG-03-94ER40833 at the University of Hawaii-Manoa.
and Grant DE-AC05-96OR22464 managed by
Lockheed Martin Energy Research Corp. at Oak Ridge National Laboratory.



\begin{thebibliography}{99}
\bibitem{EPA} Monte Carlo integration of the exact differential cross section
without assuming the equivalent photon approximation gives 2.32 rather
than 2.02 for the ratio of photon flux factors. 
This would reduce the event yield by about 15\%. Private communication,
Vladimir Savinov.
\end{thebibliography}
\end{document}